\documentclass[conference]{IEEEtran}
\IEEEoverridecommandlockouts
\usepackage{cite}
\usepackage{amsmath,amssymb,amsfonts}
\usepackage{algorithmic}
\usepackage{graphicx}
\usepackage{textcomp}
\usepackage{xcolor}
\usepackage{soul}
\def\BibTeX{{\rm B\kern-.05em{\sc i\kern-.025em b}\kern-.08em
    T\kern-.1667em\lower.7ex\hbox{E}\kern-.125emX}}

\makeatletter
\newcommand{\newlineauthors}{%
  \end{@IEEEauthorhalign}\hfill\mbox{}\par
  \mbox{}\hfill\begin{@IEEEauthorhalign}
}
\makeatother
\begin{document}

\title{Advancing Quantum Information Science Pre-College Education: The Case for Learning Sciences Collaboration}

\author{\IEEEauthorblockN{Raquel Coelho}
\IEEEauthorblockA{\textit{School of Computing and Information} \\
\textit{University of Pittsburgh}\\
Pittsburgh, United States \\
r.coelho@pitt.edu}
\and
\IEEEauthorblockN{ Roy Pea}
\IEEEauthorblockA{\textit{Graduate School of Education} \\
\textit{Stanford University}\\
Stanford, United States \\
roypea@stanford.edu}
\and
\IEEEauthorblockN{Christian Schunn}
\IEEEauthorblockA{\textit{Department of Psychology} \\
\textit{University of Pittsburgh}\\
Pittsburgh, United States\\
schunn@pitt.edu}
\newlineauthors
\IEEEauthorblockN{Jinglei Cheng}
\IEEEauthorblockA{\textit{School of Computing and Information} \\
\textit{University of Pittsburgh}\\
Pittsburgh, United States \\
jic373@pitt.edu}
\and
\IEEEauthorblockN{Junyu Liu}
\IEEEauthorblockA{\textit{School of Computing and Information} \\
\textit{University of Pittsburgh}\\
Pittsburgh, United States\\
junyuliu@pitt.edu}
\and
\IEEEauthorblockN{}  
\IEEEauthorblockA{}  
}

\maketitle

\begin{abstract}
As quantum information science advances and the need for pre-college engagement grows, a critical question remains: How can young learners be prepared to participate in a field so radically different from what they have encountered before? This paper argues that meeting this challenge will require strong interdisciplinary collaboration with the Learning Sciences (LS), a field dedicated to understanding how people learn and designing theory-guided environments to support learning. Drawing on lessons from previous STEM education efforts, we discuss two key contributions of the learning sciences to quantum information science (QIS) education. The first is design-based research, the signature methodology of learning sciences, which can inform the development, refinement, and scaling of effective QIS learning experiences. The second is a framework for reshaping how learners reason about, learn and participate in QIS practices through shifts in knowledge representations that provide new forms of engagement and associated learning. We call for a two-way partnership between quantum information science and the learning sciences, one that not only supports learning in quantum concepts and practices but also improves our understanding of how to teach and support learning in highly complex domains. We also consider potential questions involved in bridging these disciplinary communities and argue that the theoretical and practical benefits justify the effort. 
\end{abstract}

\begin{IEEEkeywords}
quantum information science, pre-college STEM education, learning sciences, interdisciplinary collaboration, design-based research, representations, educational innovation, scalable learning environments

\end{IEEEkeywords}

\section{Introduction}
Quantum mechanics has already enabled many of the defining innovations of the past century, including technologies that power the information age and the global economy as we know it \cite{b1}. Both existing technologies and emerging QIS systems embody a profound shift in scientific understanding and patterns of reasoning and problem solving \cite{b2}. They challenge us to develop new intuitions and envision possibilities that defy conventional thinking \cite{b3}. Given its vast potential, broadening early engagement with quantum information science literacy has become an urgent priority \cite{b2, b4, b5}. Achieving this goal will require attracting high school (and even younger) learners with opportunities to grapple with its fascinating core concepts, explore its unfamiliar forms of reasoning, and participate in the communities of practice critical to future innovation in a quantum age \cite{b6, b7, b8, b9, b10, b4}.

However, to make meaningful progress in this direction, we must confront two critical challenges. First, it is difficult enough to help most students grasp complex subjects like calculus or statistics \cite{b11, b12, b13, b14}, let alone the counterintuitive principles of quantum mechanics that underlie the physical systems driving emerging quantum information science technologies. Developing a deep understanding of quantum science and engineering is far from trivial. Students face demanding cognitive challenges, including high levels of abstraction, a departure from classical physics, and the difficulty of visualizing or imagining quantum systems \cite{b7, b15, b16, b17}. Second, quantum information science itself is still rapidly developing. Transitioning from supporting the learning of \emph{few} to reaching \emph{many} requires a deep inquiry about how to support learning on a scale more productively with key QIS concepts and practices.

\section{The case for a QIS-LS collaboration}
Evidence from previous STEM education efforts shows that teaching efforts not based on research have often resulted in faulty or challenged learning in complex STEM domains such as physics \cite{b18}, despite substantial federal and private sector investments in curriculum resources \cite{b19}. Even the key concepts expressed in the mass, force and motion laws of Newtonian mechanics \cite{b20, b21, b22} and something as basic as why we have seasons \cite{b23} were not remembered or understood. When cognitive and learning sciences research tackled these challenges, fundamental transformations took place in how physics was taught and in the deepening of instructionally relevant learning theories for how students come to a robust understanding of core physics concepts (e.g. \cite{b24, b25, b26}). For example, in response to these challenges, Barbara White et al. \cite{b26, b27} developed the ThinkerTools interactive software learning environment in which, by giving impulses to a virtual ball, each of which would impart a fixed velocity to the ball in a particular direction, participants engaged in activities to foster their learning of Newton’s laws of force and motion.

As Wieman and Perkins \cite{b18} observe about this history, “To move a student toward expert competence, the instructor must focus on the development of the student’s mental organizational structure by addressing the “why” and not only the “what” of the subject. These mental structures are a new element of a student’s thinking. As such, they must be constructed on the foundation of students’ prior thinking and experience of the students \cite{b28, b29}. This prior thinking may be wrong or incorrectly applied, and hence must be explicitly examined and adequately addressed before further progress is possible. The research literature on physics education can help instructors recognize and deal with particular widespread and deeply ingrained misconceptions \cite{b30, b31}. In summary, expert competence is likely to develop only if the student is actively thinking and the instructor can suitably monitor and guide that thinking. In response to these challenges of learner misconceptions, Smith, diSessa, and Roschelle \cite{b32} provide an initial sketch of a constructivist theory of learning that interprets the prior conceptions of students as resources for cognitive growth within a complex system view of knowledge, rather than conceiving a desirable pedagogical approach as exposing the misconceptions and then seeking to conquer them during instruction. 

We can learn from these STEM learning challenges of the past as we now focus on QIS. How, then, might we take STEM education to a new level by supporting broad and meaningful engagement with quantum information science?  A promising avenue lies in forging stronger connections between QIS education and learning sciences, itself an interdisciplinary field that studies how people learn and how to design environments that support learning \cite{b34, b35}. In the learning sciences, a learning environment encompasses the physical, social, psychological, and pedagogical contexts where learning occurs; it is not simply the physical space, but also the interactions, expectations, and experiences shaped by all participants – students, teachers, and staff – transactions within that learning environment \cite{b33}. As an applied discipline, learning sciences focuses on studying learning within designed environments. These environments are iteratively developed and refined to explore research questions related to learning processes, the factors that influence them, their interrelationships, and effective strategies to support both learners and learning itself \cite[p.~307]{b34}. 

The learning sciences has a well-established tradition of studying how people learn and has developed diverse and effective pathways to help learners reason productively in complex domains with emergent phenomena and system behaviors across wide orders of temporal and spatial magnitude \cite{b35, b36}. Research in the field has shown that even very young learners can engage meaningfully in STEM learning complexity with thoughtfully designed environments, representations, and supports (e.g., \cite{b37, b26}). For example, the SimCalc interactive learning environment provides Grades 2-5 students access to calculus reasoning about average speed, constant rate, and area under a rate graph by building on their kinesthetic knowledge of running, and connecting it to mathematical representations such as graphs, tables, and equations in a software model world of avatars such as creatures or elevators which move as they do in the physical world \cite{b12}. 

Learning sciences concepts, tools, and practices can offer critical resources for the emerging effort to make quantum information science not only accessible but deeply meaningful to a wider range of learners. In conjunction, quantum information science presents the learning sciences with a challenging frontier domain to determine how best to foster engaged learning and establish the desirable outcomes of understanding and applying QIS concepts, skills, and dispositions. The relationship we propose is at once reciprocal and essential: Without a deep connection between these two scientific subcultures, efforts to broaden quantum literacy risk falling short of meaningful understanding, while the learning sciences misses the opportunity to evolve its theories and methodologies in response to one of the most conceptually demanding domains of all time. Most provocatively, some reasoning strategies that QIS employs might advance how we teach and learn more broadly. QIS could lead the learning sciences field to unlock new forms of restructurations to make possible new approaches to learning. Advancing QIS literacy may thus advance human reasoning in general. 

\section{Accelerative questions for advancing QIS education with Learning Sciences collaboration}

We acknowledge the important work already underway in quantum information science education at the pre-college level (for reviews, see \cite{b38, b39, b40, b41}), which advances our understanding of how to make complex concepts relevant to quantum information systems, such as quantum states and entanglement, accessible to younger learners. However, preliminary data from our own review of existing studies on the education of quantum information science before college \cite{b42} suggest that little attention has been paid to the underlying rationale of the design and crucial considerations about learning processes. In particular, many efforts do not explicitly address how or why particular approaches, such as games that incorporate quantum phenomena into their mechanics (e.g. \cite{b43, b44, b45}), might support learning.

Although evaluating effectiveness (i.e., whether learning happened) is important, it will not be sufficient to move the field of QIS forward meaningfully. Without first asking \textit{how} and \textit{why}, we risk missing the information necessary to design and implement robust, replicable and scalable quantum education initiatives. To do this, the QIS education community must look beyond the question of 'What works?' to include 'What could work?' (which invites inquiry into mechanisms and contextual conditions) and 'What would scale?' \cite{b46}. These more demanding questions require reasoning about the specifics of both the design intervention rationale and the properties of the situated learning environments in which learning is expected to occur. The learning sciences community is particularly well placed to support QIS education in addressing these challenges through one of its signature methodological approaches: design-based research (DBR), which involves more than simply reporting results \cite[p.~152]{b47} and promotes iterative design, theory building, and evidence-based refinement in authentic learning contexts \cite{b47, b48, b49}. In addition, learning sciences offer frameworks for developing \textit{restructurations}, shifts in how learners conceptualize and engage with a discipline \cite{b50}. We expand on these ideas in the next section. 

\section{What the Learning Sciences can offer Quantum Information Science Education}

\subsection{Design-based research methodologies} Traditional controlled laboratory studies, common in previous cognitive science and education research, have provided important information but often lacked ecological validity \cite{b51}. Although these learning intervention studies can succeed under idealized conditions, they frequently fail when applied to the complex realities of real-world educational environments. Classrooms are not isolated systems. Students do not learn in a vacuum; their learning is embedded in a web of interdependent factors that vary across contexts. This includes, for example, teaching expertise, curriculum design, assessment practices, school culture, and available technologies. When researchers bring innovations that change one part of the system, they automatically change the functional organization of the whole \cite{b52, b53}.

Design-based research (DBR) \cite{b52, b54} embraces this complexity by moving from isolated interventions to iterative innovation within authentic settings. Investigating the complexity of learning is considered an integral requirement in learning sciences studies. Without tackling the complexity of studying learning in the cultural practices where learning actually happens, we lose our explanatory power and learning opportunities to design and improve learning systems \cite{b47}. We must also evolve learning theory alongside practical implementation \cite{b55}, beyond simply measuring the effects of interventions. Design-based research is basic research inspired by use \cite{b56, b57}, committed to supporting a fundamental understanding of learning while simultaneously promoting practical improvements \cite{b49}.

DBR is inspired by design sciences \cite{b58, b59, b60}, where innovation and investigation go hand in hand: Engineers build and refine innovations while simultaneously studying them in real-world contexts \cite{b52}. In education, this orientation translates to studying not only whether learning occurs, but how and why it occurs, or does not occur, under varying conditions \cite{b61}. Single, isolated experiments rarely yield generalized principles when a systemic learning environment change is involved. Sometimes learning appears to happen simply because a new tool generates initial excitement; other times, it fails due to unforeseen changes in the system or needed changes that were likewise not anticipated, such as additional support for teachers or learners themselves, such as in the One Laptop Per Child program \cite{b62}. Even when research is conducted in one naturalistic context, it will not necessarily translate to another as the context changes, so innovations also need to be studied in multiple contexts to derive useful insight into the learning process \cite{b63, b64}.

\begin{figure*}[ht]
    \centering
    \includegraphics[width=\textwidth]{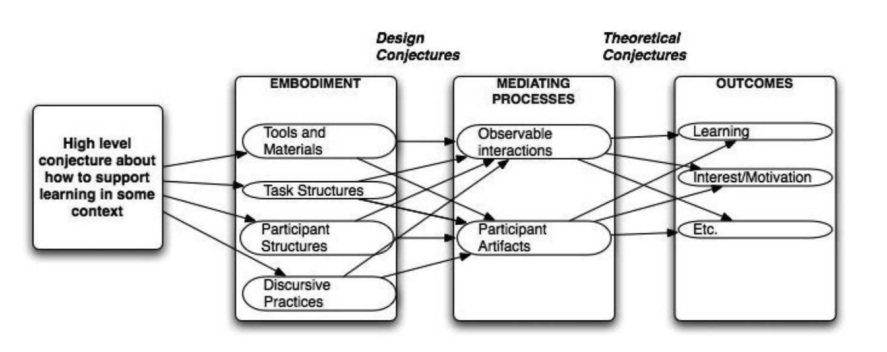}
    \caption{Generalized conjecture map for educational design research (Sandoval, 2014)}
    \label{fig:conjecture}
\end{figure*}

\begin{figure*}[ht]
    \centering
    \includegraphics[width=\textwidth]{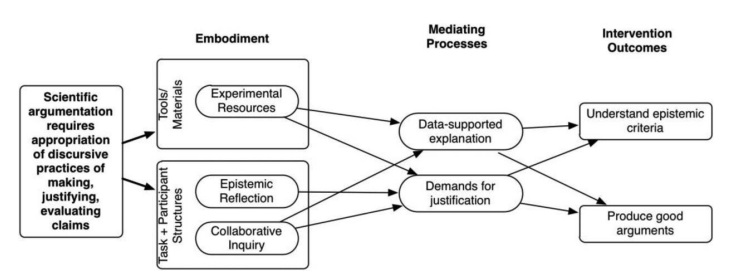}
    \caption{Initial conjecture map of a design to promote argumentation in elementary
science (Sandoval, 2014)}
    \label{fig:conjecture2}
\end{figure*}

DBR encourages multiple cycles of theory and evidence-guided learning design innovation, measuring its consequences when implemented, and cycles of refinements to improve learning processes and outcomes, helping us arrive at interventions that are both theoretically sound and practically sustainable. Articulating a learning environment design is not a side issue, but a central methodology to build robust educational solutions \cite{b65}. To make the theoretical basis of an intervention more explicit and thereby support its robustness and scalability, an influential and clearly defined operational procedure tool in the DBR tradition is 'conjecture mapping' \cite{b49}.  Conjecture maps help learning scientists articulate the relationships between high-level theoretical commitments about learning, the specific features of a designed learning environment (e.g., tools, materials, tasks, participant structures), the mediating processes that the environment is intended to support (evident through interactions or learner-produced artifacts), and the desired learning outcomes (see Fig.~\ref{fig:conjecture} for a generalized conjecture map and Fig.~\ref{fig:conjecture2} for an example in context). Conjecture maps function as a logic grammar through which learning scientists can articulate the theoretical and design conjectures embedded in a learning environment, while also guiding methodological choices to examine how these conjectures manifest in practice. The results of these tests can lead to refinements in both the design and the underlying theoretical perspective. Conjecture maps are not tools for designing learning environments, but rather for conducting research on how such environments support learning \cite[p.~20-21]{b49}. 

However, even with the strengths of DBR, there is also the issue of scalability, as scaling up educational innovation is decidedly non-trivial \cite{b66, b67, b68}. Once innovations move from one place to another without the rationales that were shared between researchers and teachers, adaptations tend to return to more familiar ways of doing things, creating what have been called 'fatal adaptations' of a learning innovation \cite{b51}. Professional learning communities \cite{b69} can help ensure that classroom teachers and school leaders understand the evidence behind the new practices they are adopting, and therefore make adaptations that are consistent with the evidence guiding their initial design.

\subsection{A methodology for developing restructurations of how to think about the discipline.}

The incremental development here is not simply a shift in the means of learning but a reconsideration of the object of learning itself. While still exploiting the benefits of design-based research, restructuration provides a transformational approach to how what counts as the object of learning is conceived. 

In the sciences of learning, "restructuring" refers to transformation in the way knowledge is organized and understood in the learner's mind. It goes beyond the acquisition of facts or skills to involve the development of new mental frameworks or the reorganization of existing knowledge structures to interpret new information and impose a new organization on what is already known \cite{b70}. These new structures then allow for new interpretations of knowledge for different accessibility to that knowledge and for changes in the interpretation and the acquisition of new knowledge. Norman \cite[p.~41]{b71} suggests that during restructuring "...we expect the learner to say, 'Oh, now I understand,' or to show evidence of jumps in understanding." Wilensky and Papert \cite{b72} describe restructuration as arising from changes in the representational infrastructure, such as the adoption of new symbolic or computational forms that alter how learners engage with and make sense of a domain. Importantly, these cognitive representational foundations develop by means of sociocultural practices \cite{b73}, such as sense making, using restructured representations, and by transformations in the participation structures for learners involved in disciplinary practices, e.g., making a scientific argument using evidence \cite{b74}. Such representational shifts and associated advances in discipline-based cultural practices do not just simplify learning; they redefine what it means to understand something, allowing for new forms of reasoning, learning, and cultural practices that were previously inaccessible.

A methodology for developing restructurations within a discipline involves designing new representational forms that provide affordances for different ways of reasoning and thinking. Affordances are the perceived and actual properties of an object that suggest how it should be used, offering clues to its function and interaction possibilities \cite{b75}. These representations are intentionally crafted because of their potential to support specific learning processes. When we begin to take the affordances of technology seriously \cite{b76}, it becomes clear that new forms of representation can fundamentally reshape the ways in which learners reason within a domain, such as linear algebra \cite{b77}.  Roschelle et al. \cite[p.~853-854]{b78} argue that 'the availability of dynamic media is changing both the practice of mathematics and the teaching of mathematics... Therefore, emerging technology changes the mathematical epistemology, that is, how people come to know, understand, and see the value of mathematical ideas'.

We find such a restructuration exemplified in research on agent-based modeling and the Logo programming environment \cite{b50}. These tools, by offering powerful representational affordances, make it easier for learners to understand and reason about complex systems. They allow users to construct and manipulate models of dynamic phenomena, thereby supporting the development of systems thinking in ways that traditional representations do not. Importantly, these representational tools are not only used in educational settings, but are also studied through a learning sciences lens. The focus is not simply on their educational deployment, but on how their design, grounded in theories of cognition and learning, can generate insights into the learning process itself. In this way, the development of new representational infrastructures becomes both a means of supporting learning and a method of advancing learning theory. In some cases, the use of these tools even reveals the need for further restructuration, prompting the design of entirely new frameworks.

A contemporary example of restructuration lies in the emerging uses of machine learning (ML) as a representational infrastructure for the sociotechnical activities of human reasoning (e.g. \cite{b79}).  In these systems, the training process becomes encapsulated in a manipulable object, the trained model. Once created, this model can be saved, shared, reused, or combined with others, effectively becoming a building block for reasoning and exploration. Without this infrastructure, engaging with certain domains would require working directly with far more complex or inaccessible mathematical representations. ML shifts this representational work by allowing interaction with the model-as-object, rather than with the algorithm or data alone. However, for ML to serve this role in learning contexts, it must also offer a user experience that enables meaningful manipulation by novices. Teachable Machine \cite{b80} is one such example. Although still limited in features and requiring further exploration from a learning sciences perspective, it begins to explore the idea of trained models as representational objects that learners can train and manipulate.

To recapitulate, another conceptual framework for what is inevitably going to be needed in bringing together quantum information science and the learning sciences lies in the way learning scientists engage with disciplinary knowledge. Using the data they collect on how learners think, reason and develop understanding, learning scientists are bound to discover alternative ways of structuring QIS knowledge for students to more effectively engage in reasoning with QIS concepts and associated competencies. The central argument then is that if we can design and embed these restructurations into curricula and learning activities, we can support deeper, more robust learning and potentially learning at a much earlier age. Restructurations might change the processes of QIS learning and even the temporal course of learning.

In the context of quantum information science, a similar opportunity emerges. Identifying key conceptual difficulties and critically examining current instructional approaches allows researchers to explore the potential for new representational forms that may serve as restructurations of the QIS domain. Establishing such innovations constitutes high-risk, high-yield research. The goal is not to simply scale up existing practices, but to fundamentally reimagine how QIS can become learnable, even for students typically considered too young to engage with such complex content. The task ahead involves more than incremental improvement; it requires a conceptual transformation, a restructuration of QIS education that may take years of iterative design, experimentation, and theoretical refinement to achieve.

It’s worth asking, then: How might we transform what is difficult in quantum information science? Specifically, what kinds of restructurations can make the core ideas of QIS, across quantum computing, communication, and sensing, accessible to youth typically considered too young to grasp such content? How might we introduce foundational concepts like complex numbers, vector spaces, or linear algebra through representational forms that do not require a complex body of prerequisite knowledge? Can students engage in reasoning within the quantum information science discipline without yet knowing the complex math behind it? What kinds of new representational infrastructure, such as agent-based models or interactive simulations, could support such a shift? How might these tools help learners connect the microlevel behavior of quantum systems to their macrolevel effects? And importantly, how might we help learners engage with the perennial question of difficult learning topics from the learner's point of view when they ask perhaps the most important question: 'So what?' Why does this matter? Why is this important to me in terms of the things that I care about?”

Insights into possible QIS restructurations are already emerging in the literature on quantum mechanics education \cite{b81}. These developments constitute a possible bridge between quantum education broadly construed and the learning sciences. For example, researchers have begun to systematically analyze how students reason with symbolic forms to make sense of quantum phenomena. This line of inquiry is already a foundational move towards restructuration, exemplifying a learning sciences approach to understanding how students construct meaning and navigate conceptual transitions within quantum mechanics. For example, a recent study \cite{b82} examines how students interpret and construct eigenvalue equations in a spins-first quantum mechanics curriculum, focusing on their reasoning as they transition from discrete to continuous systems. This shift requires not only procedural competence, but also substantial conceptual restructuring. To analyze student thinking, the study draws on multiple frameworks from the learning sciences, including symbolic forms, according to which students learn to understand physics equations in terms of a vocabulary of elements called symbolic forms, each of which associates a simple conceptual schema with a pattern of symbols in an equation \cite{b83}.

Meaningful restructuration in QIS will not result from curricular repackaging alone. It will require sustained attention to how students make sense of QIS concepts and how instruction can be designed to scaffold these understandings and more expert reasoning and representational practices through design-based research. To support this convergence, we must now consider concrete mechanisms that can bring quantum information scientists and learning scientists into sustained generative collaboration.

\section{Potential mechanisms for bringing these two communities together} 

\subsection{Creation of interdisciplinary groups within and between departments}
One promising mechanism for fostering collaboration between learning sciences and quantum information science communities is the creation of interdisciplinary groups within existing higher education institutions. These groups can originate in different parts of the university, some emerging from disciplinary departments and others from schools of education, each offering unique structural models for integration. In the former, educational research is embedded within the scientific discipline itself; in the latter, interdisciplinary training is built through interdepartmental collaboration between education and disciplinary faculty.

The University of Washington's \textit{Physics Education Group}, which was led by Lillian McDermott, offers a historical precedent for how a disciplinary department can institutionalize education-focused research and training. In her seminal Millikan Lecture, McDermott \cite{b84} drew on decades of instructional experience to critique the wave of post-Sputnik curriculum reforms, which emphasized new content and materials, but failed to adequately prepare teachers or address conceptual difficulties of students. She argued that, despite their intention to promote inquiry, these reforms were often ineffective because they relied on unrealistic expectations, such as assuming that teachers could learn content alongside their students or teach effectively using only scripted guides, and failed to consider learning theory-guided activity designs by which students actually construct an understanding of physics concepts. In response, her group adopted a research-based approach influencing the broader field of physics education that combined participation in physics and systematic investigation of how students learn, leading to instructional strategies that addressed persistent conceptual difficulties. Her group's work revealed, for example, that many students were unable to relate graphs of motion to physical phenomena, a key challenge in learning kinematics. To address this, the group designed interventions that helped students translate between real-world motion and its graphical representations. This educational research was embedded within the physics department, culminating in a Ph.D. track that allowed students to specialize in the teaching and learning of physics \cite{b85}. This structure institutionalized educational research within the discipline and created a market for such expertise across academic departments (op cit.). A parallel initiative in quantum information science could involve the formation of "quantum education groups" within physics, engineering, or computer science departments, in collaboration with schools of education. Although McDermott did not refer to learning sciences by name, likely because the field had not yet fully emerged as such \cite{b86, b87}, her work was based on cognitive science and early research on learning that would later form a foundation of the field. In addition, her emphasis on research-based instructional design places her work in clear alignment with the foundational ideas that came to define learning sciences.

Another model is the Graduate Group in Science and Mathematics Education at UC Berkeley, informally known as SESAME \cite{b88}. This interdisciplinary doctoral program was specifically designed to support the development of expertise in scientific disciplines and educational research. SESAME brought together faculty from the Graduate School of Education and a wide range of science and engineering departments, including chemistry, biology, mechanical engineering, bioengineering, computer science, and earth sciences, to support students conducting research on teaching and learning in STEM fields in both formal and informal contexts.

The program awards Ph.D.s in science, mathematics, or engineering education and requires students to reach at least a master's level competency in their discipline. SESAME students study educational theory, research methodologies, psychology, and disciplinary content and participate in teaching, seminars, and research colloquia. The structure of the program enables students to investigate a broad range of topics, including college-level STEM instruction, curriculum development for grades K-12, cognitive processes in scientific reasoning, learning technologies, and informal learning in science museums and public institutions. Although the program emphasizes educational research rather than teacher preparation, SESAME plays a crucial role in developing leaders who go on to careers in higher education, science museums, curriculum design, and industrial or nonprofit education. SESAME provides another model for quantum information science education efforts that could be emulated by forging partnerships across schools of education and disciplinary departments.

\subsection{Creating quantum information science learning centers} 
To illustrate a second mechanism by which quantum information science and learning sciences could intersect productively, we offer anecdotal evidence from one of the coauthors, who co-led a decade-long interdisciplinary collaboration at the NSF-funded multiinstitution science of learning center in the US called the LIFE Center, for Learning in Informal and Formal Environments. This initiative brought together researchers from psychology, education, computer science, neurobiology, speech and hearing sciences, neuroscience, philosophy, communication, anthropology, sociology, and media studies to study the social nature of learning. An effective strategy identified to promote interdisciplinary collaboration involved exploring and discussing empirical data from different individual investigators as a shared focal point. Co-principal investigators (co-PIs) from various projects contributed data from their own research, which then served as a common ground for discussion and exploration across disciplinary lines (e.g. \cite{b89, b90}). Graduate students, postdoctoral scholars, and faculty alike were encouraged to engage directly with the primary data of others, ask hard questions, and investigate the underlying assumptions about significance or limitations of theories and methodologies. This environment supported asking fundamental questions, which promoted deeper mutual understanding, genuine interdisciplinary learning, and, most importantly, 'reciprocal expertise affirmation', which is an essential foundation for trusting relationships among those participating in interdisciplinary investigations \cite{b91, b92}.  

For example, a neuroscientist studying cognitive processes explained the limitations of brain imaging techniques such as fMRI or MEG \cite{b93}, which limit researchers to only look at brain activity in very short bursts, sometimes seconds or minutes (fMRI) or even milliseconds (MEG). This makes it difficult to study mental processes that occur over longer periods, such as many aspects of self-regulated or social learning. Working with this neuroscientist as an interdisciplinary partner meant learning what his data could and could not reveal. His research tools gave him very precise information, but also limited what kinds of questions he could ask.

Similarly, ethnographers studying learning within the center, where the focus was on observing and analyzing multimodal data of people learning in real-world, everyday situations, offered deep, more ecologically valid \cite{b94} insights into how learning occurs in everyday life, not in scientific laboratories or schools \cite{b95}. While their work could not offer the same level of precision as brain scan data, it brought to life an essential understanding of the cultural and social dynamics that shape learning \cite{b96}. Part of working together meant acknowledging and respectfully trusting these differences and melding their respective insights into a powerful alloy stronger and more valuable than the individual components alone.

Discussions with media psychologists also brought about surprising complexity. At first, their work examining physiological responses to immersive or emotionally charged media seemed disconnected from educational concerns or real-world learning. But it became clear with more sustained collaborative research that the field has a long history of investigating how different types of media influence behavior, emotion, and cognition in everyday life, often in powerful and embodied ways. These perspectives provided additional insight into the complexity of learning experiences, particularly in digital or mediated contexts (e.g., \cite{b97}).

What these collaborations demonstrated is that joint interdisciplinary research did not emerge simply by bringing scholars together in a mandala of good intentions. It required a process of mutual discovery, learning humbly what each discipline’s tools and theories can and cannot do, building trust through shared inquiry, and eventually co-designing research that reflected the interests and strengths of all involved, certainly inspired in part by the National Science Foundation funding agency that required evidence of the value added of interdisciplinary collaboration in the Center's activities to justify renewal funding. At one stage in the ten years of the center, learning scientists, media psychologists, and neuroscientists came together to co-design and conduct studies. It brought together each of their core concerns and methodological strengths, along with their theoretical frameworks, to do work that none of them would have planned to do or executed on their own. Without this collaboration, the disciplinary boundaries likely would have remained intact, and researchers may not have felt compelled to articulate the limitations of their approaches to scholars in another discipline. Importantly, this work was not only intellectually rewarding; it was structurally incentivized. As an NSF program officer noted, funding depended on demonstrating that the research being conducted was truly interdisciplinary, that is, work that would not have happened within programs of research for individual disciplinary silos.

LIFE illustrates that a QIS-LS center collaboration could start, for example, with shared data exploration. On the QIS side, researchers and educators might bring examples of student interactions and reasoning about quantum devices, simulations, or problem-solving strategies to the collaboration. Learning scientists, in turn, could offer perspectives on how they study and support learning in complex, real-world settings \cite{b52, b98}. 

For QIS and Learning Sciences, both the interdisciplinary degree programs and associated faculty collaborations, and the LIFE Center experience suggest promising pathways toward robust collaboration.

\section{Questioning and reaffirming the case for QIS-LS collaboration}

\subsection{Can't we simply leverage existing STEM education research?}
One might wonder whether interdisciplinary inquiry at the boundary between quantum information science and the learning sciences is necessary, arguing instead that one might simply rely on existing knowledge from discipline-based education research communities in physics, chemistry, and computer science, fields that already engage with learning sciences in meaningful ways. Although we agree that previous findings should be leveraged, these advances must be seen as a foundation, not a substitute, for further inquiries. Research in other STEM domains does not magically translate into effective strategies for teaching and learning for quantum information science, a domain with its own conceptual frameworks and learning challenges. Although efforts to improve quantum mechanics education have attracted growing attention through research in physics education \cite{b81}, QIS introduces additional layers of complexity. QIS requires learners to not only engage with but also reason using abstract and often non-intuitive principles through the lenses of computation and information theory \cite{b99}. QIS requires new interdisciplinary modes of reasoning. For example, how does quantum computational thinking differ from classical computational thinking \cite{b100}? What does it mean to think and engage in argumentation algorithmically within a quantum paradigm, where uncertainty and parallelism are intrinsic features? How do entanglement and superposition enable entirely new technological capabilities? These are not only technical questions, but profoundly educational ones. Moreover, designing quantum algorithms is enormously complex, and the field itself is still developing to understand what kinds of problems quantum computers may ultimately solve more efficiently than classical ones \cite{b99}. If this is where the field needs to advance, then we must ask: What are the first conceptual foundations, intuitions, representations, and practices that learners need to eventually participate in the social and cultural practices of quantum algorithm design? What kinds of learning experiences are best for supporting the development of these fundamental ways of thinking? These are not questions that disciplinary research alone can answer. Instead, they require intentional, interdisciplinary collaboration between QIS experts and learning scientists, drawing on insights from discipline-based education research in physics, chemistry, and computer science, to design, study, and refine learning environments that move beyond classical assumptions and support learners in navigating this emerging computational paradigm.

\subsection{What about learning engineering?}
Furthermore, although the emerging field of \textit{learning engineering} \cite{b34, b101}  may appear to offer a promising pathway to designing and scaling educational solutions of quantum information science, it is crucial to distinguish its aims from those of learning sciences. Learning engineering is a professional practice. Learning engineers, by their own descriptions, apply insights from the learning sciences (and related disciplines) to systematically design, evaluate, and refine educational solutions, such as technologies, curricula, or infrastructures, for particular populations with specific learning goals. Their goal is not to advance fundamental understanding, but to deliver effective, scalable, and context-sensitive solutions \cite[p.~308]{b34}. Learning engineering represents pure applied research \cite{b57}. Although theoretical refinement may occur as a byproduct of work conducted by members of the learning engineering community, it is not a central aim of their research.

Learning engineering efforts often emerge from use-inspired basic research initiatives once promising prototypes, tools, or pedagogical models show potential for broader effects. Scaling such innovations into widely usable educational products requires not only applying insights from the learning sciences, but also involving multidisciplinary teams, including software engineers, instructional designers, data analysts, user experience designers, and domain experts. For example, Scratch, a visual block-based programming language originally developed in Mitchel Resnick’s research lab at MIT, began as an approach to support programming through personally meaningful creative expression, such as interactive stories and games that could be shared with others \cite{b102, b103}. Only after years of research and refinement was it spun out into the Scratch Foundation to support its widespread adoption in vast global online communities \cite{b34}. 

Regardless of ongoing debates between those who see little distinction between learning engineering and learning sciences \cite{b101} and those who view them as fundamentally different but complementary enterprises \cite{b34}, it is the use-inspired basic research conducted within the learning sciences, not solution-oriented applied research of learning engineering, that the quantum information science education currently needs at the pre-college level.  This is because QIS, as an emerging and conceptually challenging domain, demands not only improvement in practice but a deeper understanding of how learners engage with counterintuitive ideas and how best to support their conceptual development over time. Moving from supporting the learning of a few to reaching the many will require more than scaling up access to quantum computational tools and platforms. This ambitious goal demands the design and systematic study of innovative learning environments tailored to the unique challenges of this emerging domain, as well as foundational insights into the cognitive, affective, social, and cultural dimensions of learning in quantum information science. The learning sciences, with its simultaneous threefold commitment to a) designing innovative learning environments, b) generating knowledge about how these environments function in their intended contexts, and c) advancing fundamental understanding of learning or teaching \cite[p.~19]{b49}, offers a more appropriate foundation for improving both theory and practice in quantum information science education. The latter focus on foundational learning lies outside the scope of learning engineering.

\subsection{Isn't Interdisciplinary collaboration hard to do?}
Although increased collaboration between quantum information science and learning sciences holds promise, scholarship on interdisciplinary research \cite{b104, b105} urges caution about how such partnerships are initiated and implemented. Interdisciplinary work is inherently complex and challenging. Funded interdisciplinary initiatives, while offering higher levels of financial support, come with increased competition and administrative burden. The assembly of diverse teams can dilute funding per PI, and managing interdisciplinary collaborations often involves navigating differences in status, publication norms, and disciplinary expectations. Physical separation between collaborators, along with the need to build a shared and respectful understanding between fields, further complicates these efforts. University-based programs face their own set of challenges. Institutional structures often lack the cultural and infrastructural support to sustain interdisciplinary scholarship. Faculty, particularly early-career researchers, can find themselves isolated from their disciplinary homes and unsupported in promotion and evaluation systems that continue to favor traditional, discipline-specific contributions. In addition, interdisciplinary programs are frequently underfunded and vulnerable to closure. 

Despite these challenges, the benefits of interdisciplinary research are well documented. The evidence overwhelmingly suggests that boundary-spanning teams have better results, including greater productivity and scientific influence, compared to less distributed teams or solo scientists \cite[p.~7]{b106}. Furthermore, cross-disciplinary teams produce more publications and publish in more diverse publication venues \cite{b107, b108, b109}, and generate more innovative products than comparison teams \cite{b110, b111, b112, b113}. 

To realize these benefits within QIS-LS, more is required than simply bringing the two fields into contact; it requires deeper integration to avoid reproducing the well-documented challenges of cross-disciplinary work \cite{b114}. Cultural alignment and infrastructure support are necessary prerequisites. However, what repeatedly allows genuine collaboration to take root is the cultivation of shared understanding, respect for different forms of expertise, and sustained investigative interactions around a common problem space \cite{b106}. The LIFE Center funded by the NSF exemplifies this process, illustrating how interdisciplinary collaboration can thrive when rooted in the collective examination of empirical data and a deep respect for diverse forms of expertise. These are the very conditions that Galison \cite{b115} described as constituting a trading zone, an intersection where disciplinary differences are used to address shared problems in ways that no single field could achieve alone.

\section{Aspirational Prospects for Pre-college Quantum Information Science}
The importance of pre-college quantum information science education should not be underestimated. As we enter a new technological era driven by quantum computing, communication, and sensing, our education systems need to evolve to meet the new demands. We can foresee QIS learning environments that reflect the collaborative, inquiry-driven, and problem-centered nature of actual scientific practices \cite{b116}. 

This vision is illustrated in the CoVis Project (e.g., \cite{b117, b118}), which over five years created a model in which geographically dispersed students, teachers, and scientists co-designed, experimented, and learned together through a sociotechnical platform of collaborative visualization (CoVis). Rather than simply using advanced tools and networks for teachers to ‘deliver’ textbook information and students to ‘consume’ knowledge, but promoting building knowledge activities, CoVis found that when students and teachers have access to powerful scientific tools, such as visualization software, real-time data streams, structured groupware, and telementors, they begin to think and operate more like scientists, creating conditions for designing computer-supported collaborative learning technological innovations to better serve their emerging needs \cite{b119}. They engage in authentic inquiry, collaborate across distances, and contribute to the ongoing design and refinement of their learning tools. Analogously, to support QIS education, new platforms could allow students not only to learn quantum concepts but also to simulate quantum systems, build quantum algorithms, collaborate on entanglement experiments, and engage in real-time discussions with researchers and graduate students at the frontiers of the discipline.

\section{Conclusion}
The need to broaden early engagement with quantum information science requires more than simply exposing students to quantum concepts; it calls for thoughtfully designed, theoretically grounded learning environments that support deep understanding of ideas that challenge even expert intuition. A promising path forward lies in sustained collaboration between QIS and the learning sciences, which bring powerful theoretical frameworks and methodologies, such as design-based research, to bear on the creation of robust, scalable, and context-appropriate learning environments. While interdisciplinary collaboration is never easy, its potential is greatly significant: not only could it make quantum ideas and practices more accessible to younger learners, but it can also reciprocally enrich the learning sciences by pushing the boundaries of how we understand knowledge-building for unfamiliar conceptual domains. We aim to explore how engaging with QIS can open new directions for the learning sciences in future work.

A promising starting point for fostering this collaboration is the organization of a transdisciplinary workshop that brings together researchers from quantum information science and the learning sciences. Although global efforts have begun to identify \textit{what}, that is, the key QIS concepts to ground K-12 curricula and educational initiatives (e.g. \cite{b120}), it is necessary to also address the \textit{why} and \textit{how}. The primary objective of this workshop would be to identify preeminent research directions that can shape the future of pre-college quantum information science education. Given ongoing discussions surrounding the establishment of national quantum education centers (e.g. \cite{b121}), the current moment presents a particularly opportune time to initiate such collaborative work.

\end{document}